\documentclass{sig-alternate.fix-ref-space}

\usepackage{xspace}

\newcommand{\omt}[1]{}

\newcommand{\query}{q}
\newcommand{\origQuery}{q^*}
\newcommand{\QuerySet}{Q}
\newcommand{\doc}{d}

\newcommand{\docindex}{i}

\newcommand{\corpus}{{\cal D}}
\newcommand{\word}{w}
\newcommand{\wordseqlength}{n}
\newcommand{\wordseq}{\word_1 \word_2 \cdots \word_\wordseqlength}

\newcommand{\wordseqvar}{x}
\newcommand{\arbNumber}{k} 
\newcommand{\freq}[2]{{\rm tf}(#1 \in #2)}
\newcommand{\wordIndex}{j}

\newcommand{\set}[1]{\{#1\}}
\newcommand{\definedas}{\stackrel{def}{=}}

\newcommand{\kld}[2]{D\left(#1 \,\,\big\vert\big\vert\,\, #2\right)}

\newtheorem{definition}{Definition}

\newcommand{\parmark}{\hspace*{.1in}} 

\newcommand{\clust}{c}

\newcommand{\clusters}[1]{{\cal C}} 
\newcommand{\clusterswith}[1]{\clusters{}(#1)}

\newcommand{\sizeclust}{\gamma} 

\newcommand{\arbGroup}{X}
\newcommand{\arbGroupMember}{x}
\newcommand{\altArbGroupMember}{y}

\newcommand{\prob}{p}

\newcommand{\ilmprob}{\prob}

\newcommand{\inducedprob}[3]{\ensuremath{#1_{#2}(#3)}}

\newcommand{\mlprob}[2]{\inducedprob{\ilmprob^{ML}}{#1}{#2}}

\newcommand{\renderprob}[2]{\inducedprob{\ilmprob}{#1}{#2}} 
\newcommand{\dirichletParam}{\mu} 
\newcommand{\dirichletLM}[3]{\inducedprob{\ilmprob^{[\dirichletParam]}}{#1}{#2}}
\newcommand{\bareMLProb}[1]{\ilmprob^{ML}_{#1}}
\newcommand{\bareDirProb}[1]{\ilmprob^{[\dirichletParam]}_{#1}}

\newcommand{\numretdocs}{N}

\newcommand{\firstmention}[1]{{\bf #1}}
\newcommand{\baseAlgName}{Audition}
\newcommand{\mcDocToDoc}{{\mcfull}Doc-\baseAlgName\xspace}
\newcommand{\fifoAlg}{Viterbi \mcDocToDoc}
\newcommand{\mcfull}{} 
\newcommand{\mc}{}
\newcommand{\mcClusterToDoc}{{\mcfull}Cluster-\baseAlgName\xspace}

\newcommand{\rp}{$\mbox{P}^*$\xspace}
\newcommand{\mixtureAlg}{\rp Interpolation\xspace}
\newcommand{\iterMixAlg}{Iterated \mixtureAlg}
\newcommand{\boosting}{Iterated Truncation\xspace}
\newcommand{\rerank}{Truncated \rp Re-rank\xspace}
\newcommand{\rerankBoosting}{Iterated {\rerank}\xspace}
\newcommand{\mixtureParam}{\lambda}

\newcommand{\methodArb}{M}

\newcommand{\abbrevFifo}{{\rm VDoc}}
\newcommand{\abbrevMCdocToDoc} {\rm {{\mc}Doc}}

\newcommand{\abbrevMCclustToDoc} {\rm {{\mc}Clust}}
\newcommand{\abbrevMixtureAlgo}{\mixtureAlg}
\newcommand{\abbrevNoPrev}{none}
\newcommand{\abbrevIntRer}{\abbrevMixtureAlgo}
\newcommand{\abbrevIterIntRer}{Iterated \abbrevMixtureAlgo}
\newcommand{\abbrevMixPrefix}{Int-\xspace}
\newcommand{\abbrevMixMCclustToDoc} {\rm {\abbrevMixPrefix}{\abbrevMCclustToDoc}}
\newcommand{\abbrevMixMCdocToDoc} {\rm {\abbrevMixPrefix}{\abbrevMCdocToDoc}}
\newcommand{\abbrevReRank}{\rerank}
\newcommand{\abbrevBoosting}{\boosting}
\newcommand{\abbrevRerankBoosting}{\rerankBoosting}
\newcommand{\abbrevItFifo}{\abbrevFifo}

\newcommand{\render}{render\xspace}
\newcommand{\renderer}{renderer\xspace} 
\newcommand{\Renderer}{Renderer\xspace} 
\newcommand{\renderers}{{\renderer}s\xspace}
\newcommand{\Renderers}{{\Renderer}s\xspace}
\newcommand{\Rendition}{Rendition\xspace}
\newcommand{\rendition}{rendition\xspace}
\newcommand{\renditions}{{\rendition}s\xspace}
\newcommand{\renders}{renders\xspace}

\newcommand{\pfDocText}{pseudo-query\xspace}
\newcommand{\pfDocTextCapital}{Pseudo-Query\xspace}

\newcommand{\pfDocsText}{pseudo-queries\xspace}

\newcommand{\PfDocsText}{Pseudo-queries\xspace}
\newcommand{\rendererVar}{r}
\newcommand{\rendererSet}{{R}}
\newcommand{\genDoc}{\doc} 
 
\newcommand{\score}{{\rm Score}}

\newcommand{\stepT}{t}
\newcommand{\maxStepT}{\rho}

\newcommand{\algScoreAtStep}[3]{\score_{#1}(#2)} 

\newcommand{\algScoreAtStepT}[2]{\algScoreAtStep{#1}{#2}{}}

\newcommand{\scoreByMethod}[2]{\score_{#2}(#1)}
\newcommand{\scoreByMethodFinal}[2]{\score^{(\maxStepT)}_{#2}(#1)}

\newcommand{\normVar}{K}
\newcommand{\normfactor}[2]{\normVar(#1; #2)}
\newcommand{\normfactorDocToClust}[1]{\normVar_1(#1)}
\newcommand{\normfactorClustToDoc}[1]{\normVar_2(#1)}
\newcommand{\pfIndicator}[1]{\widehat{#1}}
\newcommand{\pfDoc}{\pfIndicator{\query}}
\newcommand{\pfDocs}{\pfIndicator{\QuerySet}}
\newcommand{\pfWeightText}{weight\xspace}
\newcommand{\pfWeightsText}{{\pfWeightText}s\xspace}
\newcommand{\pfWeightedText}{{\pfWeightText}ed\xspace}
\newcommand{\pfScore}{\pfIndicator{w}}

\newcommand{\maxRankDocIndex}[1]{i^+} 
\newcommand{\maxRankDoc}{\pfDoc^+}
\newcommand{\repertoireText}{repertoire\xspace}
\newcommand{\repertoiresText}{{\repertoireText}s\xspace}
\newcommand{\abbrevRep}{{\rm Rep}}
\newcommand{\repertoire}[4]{\abbrevRep(#1 ; #2\, \vert\, #3, #4)}
\newcommand{\repertoireBareArb}[2]{\abbrevRep(\rendererVar)}
\newcommand{\repertoireFullArb}{\repertoire{\rendererVar}{\arbGroup}{\rendererSet}{\arbNumber}}
\newcommand{\bestDocToDoc}{\abbrevRep({\genDoc}\vert\spreadDocToDoc)}
\newcommand{\bestDocToClust}{\abbrevRep({\clust}\vert\spreadClustToDoc)}
\newcommand{\bestClustToDoc}{\abbrevRep({\doc}\vert\spreadDocToClust)}

\newcommand{\overgeneralCap}{Query drift\xspace}
\newcommand{\overgeneralCAP}{Query Drift\xspace}
\newcommand{\overgeneralLow}{query drift\xspace}
\newcommand{\overgeneralLowH}{query-drift\xspace}

\newcommand{\env}{{\rm TopRen}} 
  
\newcommand{\spreadClustToDoc} {\tau} 
\newcommand{\spreadDocToClust} {\sigma} %

\newcommand{\spreadDocToDoc} {\tau} 
\newcommand{\pfNum}{\spreadDocToDoc_1}
\newcommand{\numNormDocToDoc}{m}

\newcommand{\envformat}[6]{#5(#3; #2, #4)} 
\newcommand{\envOfElement}[4]{\envformat{#1}{#2}{#3}{#4}{\env}{\in}}
\newcommand{\envOfText}[3]{\envOfElement{#1}{#2}{\wordseqvar}{#3}}

\newcommand{\groupArbEnvOfText}[1]{\envOfText{\rendererVar}{#1}{\arbNumber}}

\newcommand{\relModel}{{\cal R}}

\newcommand{\relModelText}{{relevance model}\xspace}
\newcommand{\clipped}{clipped\xspace}
\newcommand{\clipping}{clipping\xspace}

\newcommand{\abbrevRelModel} {RelModel}
\newcommand{\abbrevRelModelClipped} {ClippedRelModel}

\newcommand{\rocDiff}{\eta} 
\newcommand{\relDiff}{{\cal R}} 
\newcommand{\clippedDiff}{c}

\numberofauthors{3}

\title{Better than the Real Thing? Iterative \pfDocTextCapital Processing using
  Cluster-Based Language Models}

\begin{document}
\conferenceinfo{SIGIR'05,}{August 15--19, 2005, Salvador, Brazil.}
\CopyrightYear{2005}
\crdata{1-59593-034-5/05/0008}

\author{
\alignauthor Oren Kurland $^{1,3}$ \\
\email{kurland@cs.cornell.edu} \\[5mm]
\begin{tabular}{l}
\affaddr{1.~Computer Science Department,
Cornell University, Ithaca NY 14853, U.S.A.}  \\
\affaddr{2.~Language Technologies Institute, Carnegie
Mellon University, Pittsburgh PA 15213, U.S.A.} \\
\affaddr{3.~Computer Science Department,
Carnegie Mellon University, Pittsburgh PA 15213, U.S.A.} \\
\affaddr{4.~William Davidson Faculty of
Industrial Engineering and Management, Technion, Haifa 32000, Israel}
\end{tabular}\\
\alignauthor Lillian Lee $^{1,2,3}$ \\
\email{llee@cs.cornell.edu}
\alignauthor Carmel Domshlak $^{4}$
\email{dcarmel@ie.technion.ac.il}
}

\maketitle

\begin{abstract}
We present a novel approach to pseudo-feedback-based ad hoc retrieval
that 
uses language models induced from both documents and clusters.  
First, we treat the pseudo-feedback
documents
produced in response to the original query as a
set of {\em \pfDocsText} that themselves can serve as  input to the
retrieval process.  
Observing that 
the documents returned in response to the  \pfDocsText can
then act as \pfDocsText for 
subsequent rounds, we 
arrive at a
formulation of \pfDocText-based retrieval as an iterative process.
Experiments show that several concrete instantiations of this idea,
when applied in conjunction with techniques designed to
heighten precision,  yield performance results rivaling those of a number of
previously-proposed algorithms, including the
standard language-modeling approach.  The use of
{\em cluster-based} language models is a key contributing factor
to our algorithms' success.

\end{abstract}

\vspace{1mm}
\noindent
{\bf Categories and Subject Descriptors:} H.3.3 {[Information Search
and Retrieval]}: {Retrieval models, Clustering}

\vspace{1mm}
\noindent
{\bf General Terms:} Algorithms, Experimentation

\vspace{1mm}
\noindent
{\bf Keywords:}  language modeling, clustering, pseudo-feedback, pseudo-queries, rendition,
query drift, cluster-based language models, aspect recall

\section{Introduction}
\label{sec:intro}
Statistical language models have become an important tool in
information retrieval, 
and have been applied to many settings
\cite{Croft+Lafferty:03a}. 
In the 
case of fully automatic
ad hoc IR, where the task is to find documents relevant to a query $\query$
without access to 
relevance-feedback information,  a great
deal of recent research builds upon Ponte and Croft's
initial proposal \cite{Ponte+Croft:98a} wherein the rank of a
document $\doc$ is based on the probability 
assigned to $\query$ by a language model constructed from $\doc$.
We can gloss this
ranking principle as, ``retrieve the documents that are the best
{\em \renderers} of the query''.\footnote{Our choice of terminology
--- ``\renderers'' rather than ``generators'' ---
reflects the fact that we do not assume that 
documents (or their
induced language models) are the source that ``generates'' $\query$: A monkey
randomly striking typewriter keys may produce a word-for-word copy of
{\em Hamlet}, but we do not therefore say that it is the author of the
play.}

The work presented in this paper is partly motivated by the following
hypothesis:   documents that are the best {\em \renderers} of a query may
be good alternate {\em \renditions} of it.  
Indeed, a basic premise
behind query-expansion techniques utilizing pseudo-feedback is that 
top-retrieved documents may reveal dimensions of the user's
information need that are not obvious from the original (short) query
\cite{Ruthven+Lalmas:03a}.\footnote{This is another motivation
behind our terminology: in the hands of a skilled artist, a rendition
of a particular piece may be faithful to the original in many
respects, and yet still be superior overall.} Assuming for now that the hypothesis is
true (we discuss it further below), we therefore propose a type of
pseudo-feedback approach in which 
query ``expansion'' consists of wholesale replacement of $\query$
with a list of {\em \pfDocsText} consisting of
the query's best \renderers.  

\PfDocsText are clearly a form of pseudo-feedback.
However, the former term suggests that once we have created
\pfDocsText from the initial {query}, we can in principle repeat the
process, this time seeking the top \renderers of the {\pfDocsText}.
And if the \pfDocsText are indeed more informative than their predecessor(s),
then we expect this repetition to improve the retrieval results.
We thus arrive at an
{iterative} {\em boot-strapping} approach in which the
previously-retrieved best \renderers become the \pfDocsText for the
next round.

Unfortunately, pseudo-feedback quality 
can suffer from problems with both
precision and ``aspect recall''.  The 
currently unavoidable phenomenon
of non-relevant documents appearing 
in the retrieval results
leads to
{\em \overgeneralLow}, ``the alteration of the focus of a search topic
caused by improper expansion''
\cite{Mitra+Singhal+Buckley:98a}\footnote{We are not referring to
  ``query drift'' in the sense of user interests changing over time
\cite{Allan:96a}.}.  
As for recall, key aspects of the user's information need may be
completely missing from the pool of top-retrieved documents, due to
both small pool size (in order to keep precision reasonable) 
and selection for documents most resembling the short --- and hence
potentially not completely informative ---  initial query.
The inability to cope with
this missing-aspect problem is
viewed as
 a
major failing of current
systems \cite{Buckley:04a,Harman+Buckley:04a}.

To increase aspect recall, we use the structure of the corpus,
as manifested through language models built on 
document clusters, 
to suggest and represent potential facets of the
user's needs
\cite{Kurland+Lee:04a,Liu+Croft:04a}.  In particular, we consider
finding good \renderers among a set of {\em clusters} rather than the set
of documents: 
a cluster that is a good \renderer may contain
documents that,
while relevant,  superficially 
don't match the
query string precisely because they
include aspects not immediately evident in $\query$. (Such documents
could be present in the cluster by dint of being similar to other
relevant documents 
with respect to non-query terms.)

As for 
query drift,
the problem
would seem to be exacerbated by the multiple
iterations performed by our algorithm, since poor-quality input early
in the pipeline has a potentially disastrous effect on retrieval
results later on. 
To cope with this difficulty, we provide a number of methods  
that ``re-anchor''
\pfDocsText to the original query.

Experiments with several large corpora reveal significant improvements
in both average precision and recall
over the standard
language-modeling approach
(which corresponds to a degenerate version of our methods).
This finding suggests that
\pfDocsText 
may 
indeed 
be
better than the
original query as a basis for retrieval.
Moreover, comparisons
against two highly effective techniques incorporating pseudo-relevance
feedback --- Rocchio 
on pseudo-feedback and Lavrenko and Croft's
language-model-based {\em relevance model}~\cite{Lavrenko+Croft:01a} 
--- show that our cluster-based methods
can often
 provide competitive
or superior performance.

\section{Retrieval Framework}
\label{sec:retMethods}

We now present a suite of fully automatic iterative
algorithms for processing \pfDocsText.  
As mentioned above, all our algorithms conform to 
the same
 general format:
find the best \renderers of a current set of \pfDocsText; then,
repeat the 
process using these best \renderers as the new \pfDocsText.

We begin by establishing some notation and conventions in
Section \ref{sec:notation}.  Then,  we  discuss the two main axes along which our algorithms
vary.  The first such axis, detailed in Section \ref{sec:tracers},  is
the basic definition of a good 
\renderer; the options we consider are:
 (1) a document that is a good \renderer of at least {one}
\pfDocText; (2) a document that is a good  \renderer of {multiple}
\pfDocsText; and (3) a {cluster} 
that is a good  \renderer of multiple
\pfDocsText.  The second main axis of algorithm variation, discussed in Section \ref{sec:iterate}, is the choice of  mechanism for
preventing \overgeneralLow.  One idea we pursue is to incorporate
the \rendition probability assigned to the original query.

\subsection{Notation and Conventions}
\label{sec:notation}

Throughout this section, $\corpus$ denotes a given document set, which
induces a fixed vocabulary.
The notation $\origQuery$ indicates the user's initial query, and $\numretdocs$
stands for the number of documents to be returned in response to
$\origQuery$ when the retrieval process terminates.   
Lower-case Greek
letters indicate algorithm parameters that were varied in our experiments.

We use
$\clusters{\corpus}$ to refer to a set of clusters of the documents in
$\corpus$ (in our work, $\clusters{\corpus}$ is computed prior to
retrieval time).  We freely switch between thinking of a cluster as a
subset of $\corpus$ and thinking of it as the single text string created
by concatenating its constituent documents in some pre-defined
order
\footnote{In our work, the concatenation order is irrelevant since we use unigram language models.
}
--- this allows us to treat queries, documents, and clusters uniformly
as sequences of terms.

The algorithms
we present engage in iterative processing of \pfDocsText.
In what follows,
we assume that in each of the
$\maxStepT$ rounds, the \pfDocText input 
consists of a
ranked 
list $\pfDocs = \pfDoc_1, \pfDoc_2, \ldots$ together with 
a
{\em \pfWeightText} function $\pfScore: \pfDocs \rightarrow [0,1]$ such that 
$\pfScore(\pfDoc_\docindex) \geq \pfScore(\pfDoc_{\docindex + 1})$.
In the first iteration, $\pfDocs = \origQuery$, and we set
$\pfScore(\origQuery)$ to be $1$.  In subsequent iterations,
$\pfDocs$ is an ordering of the documents in $\corpus$.

A key concept in our work is that of a \renderer $\rendererVar$'s {\em
\repertoireText} among a set of text strings, by which we mean the subset of
the strings that $\rendererVar$ is a {\em top \renderer} of.
We therefore make the following  definitions.  Let
$\renderprob{\rendererVar}{\wordseqvar}$ denote an estimate of the
probability that $\rendererVar$ (in our work, either a document or
a cluster) \renders the text
sequence $\wordseqvar$.  
\begin{definition}
\label{defn:docEnv}
Let $\wordseqvar$ be a text sequence, and let $\rendererSet$ be a 
finite 
set of potential \renderers.
Then, $\wordseqvar$'s  {\em top $\arbNumber$ \renderers in
$\rendererSet$}, denoted 
$\groupArbEnvOfText{\rendererSet}$,
is the set of $\arbNumber$ 
items
$\rendererVar \in \rendererSet$ that yield the
highest
\footnote{Throughout
this paper, we assume that
the corpus documents and clusters are identified by numeric labels,
  with ties 
broken in favor of lower-numbered items.}  
$\renderprob{\rendererVar}{\wordseqvar}$.
\end{definition}
\begin{definition}
For \renderer $\rendererVar$, set of text strings $\arbGroup$,
set of potential \renderers $\rendererSet \ni \rendererVar$,  and  positive integer $\arbNumber$,
we define the {\em
  \repertoireText of $\rendererVar$ 
in
  $\arbGroup$ 
with respect to
$\rendererSet$ and $\arbNumber$}  as
$$\repertoireFullArb \definedas \set{\arbGroupMember \in \arbGroup : \rendererVar \in
\envOfText{\rendererVar'}{\rendererSet}{\arbNumber} 
}.
$$
For compactness, we suppress $\arbGroup$, $\rendererSet$, and/or $\arbNumber$ 
in our
notation when no confusion can result.
\end{definition}
{Note that {\repertoireText}s are sets, not sorted lists; thus, when
we say that a \pfDocText $\pfDoc \in \repertoireBareArb{\rendererVar}{\arbNumber}$ is ``highly
ranked'', 
we mean that it occurs early in $\pfDocs$, as opposed to,
say, that its \rendition probability
$\renderprob{\rendererVar}{\pfDoc}$ is large
with respect to the other members of $\repertoireBareArb{\rendererVar}{\arbNumber}$.}

\subsection{Basic Methods for Scoring \Renderers}
\label{sec:tracers}

We now present three basic options for 
determining
the best \renderers
of a given iteration's \pfDocsText.
Each such method $\methodArb$ takes as input the ranked list of
\pfDocsText $\pfDocs$ and the associated \pfDocText \pfWeightsText
$\pfScore(\pfDoc_\docindex)$ and  produces a score
$\scoreByMethod{\doc}{\methodArb}$ for each document $\doc$.  The input to the
next round can then be created by setting $ \pfScore(\doc) =
\scoreByMethod{\doc}{\methodArb}$ and sorting all the documents in
$\corpus$ by this quantity, unless additional mechanisms for coping
with \overgeneralLow are applied (see Section \ref{sec:iterate}).

We begin by
restricting our consideration of possible \renderers to documents. 
The \firstmention{\fifoAlg} scoring method is a straightforward
procedure that  ranks those documents 
with \repertoiresText containing a highly-\pfWeightedText \pfDocText above those
that are top \renderers only of lower-\pfWeightedText ones.
Specifically, in each round, it first returns the
top $\spreadDocToDoc$ 
\renderers of $\pfDoc_1$, then the top
$\spreadDocToDoc$ 
\renderers of $\pfDoc_2$
(repeated \renderers discarded),
and so on, with the
list of
top \renderers $\genDoc$ of each $\pfDoc_\docindex$
sorted in descending order of 
$\renderprob{\genDoc}{\pfDoc_\docindex}$.\footnote{
Although it does not convey more insight than
the English description just given, for the sake of completeness, 
here is 
a formulation 
of the \fifoAlg scoring method 
in terms of an explicit 
score
function.
 For a given document 
$\genDoc \in \corpus$, let 
$\maxRankDoc$ be the highest-ranked \pfDocText $\pfDoc$ in
$\bestDocToDoc$, and let
$\maxRankDocIndex{\doc}$ be $\maxRankDoc$'s rank.
(If $\genDoc$ is not a top \renderer of any \pfDocText, we set $\maxRankDocIndex{\doc}$ to $|\corpus|+1$ and
$\maxRankDoc$ to the dummy value $\origQuery$.) Then, 
we define
\begin{equation*}
\algScoreAtStepT{\abbrevFifo}{\genDoc} \definedas
\frac{ \renderprob{\genDoc}{\maxRankDoc} + 2(|\corpus| -
 \maxRankDocIndex{\genDoc} + 1)}
{1 + 2|\corpus|}.
\end{equation*}
}
Hence,
suppose
we have two top \renderers of $\pfDoc_1$, $\genDoc$ and $\genDoc'$,
such that $\renderprob{\genDoc}{\pfDoc_1} >
\renderprob{\genDoc'}{\pfDoc_1}$.
Then  $\genDoc$ would be ranked above
$\genDoc'$ even if $\renderprob{\genDoc}{\pfDoc_\docindex} \ll \renderprob{\genDoc'}{\pfDoc_\docindex}$ for every other $\pfDoc_\docindex$.

If we were confident that $\pfDoc_1$ is indeed the 
best
representation of the user's information need, then such behavior is not
unreasonable.  
In practice, however, 
such confidence may not be warranted; rather,
we might 
consider a document that \renders
many of the \pfDocsText to be potentially as good or
better a candidate for retrieval than a document that \renders only
one.
Hence, we define an alternative scoring method,
\firstmention{\mcDocToDoc}. 
In essence, it
rewards a potential \renderer $\genDoc \in \corpus$ for every
\pfDocText in its \repertoireText, although less credit is assigned for 
low-\pfWeightText 
$\pfDoc_\docindex$
and for $\pfDoc_\docindex$ that $\genDoc$ assigns a low \rendition
 probability to. 
 Specifically, ranks are induced by 
the following function:
\begin{equation}
\label{eq:mcDocToDoc}
\algScoreAtStepT{\abbrevMCdocToDoc}{\genDoc} \definedas  
 \sum_{\pfDoc \in \bestDocToDoc}
\pfScore(\pfDoc) \cdot \frac{\renderprob{\genDoc}{\pfDoc}}{\normfactor{\pfDoc}{\numNormDocToDoc}},
\end{equation}
where the re-scaling term $\normfactor{\pfDoc}{\numNormDocToDoc} = \sum_{\doc' \in
\envOfElement{a}{\corpus}{\pfDoc}{\numNormDocToDoc}     }
 \renderprob{\doc'}{\pfDoc}$ serves to 
compare
$\renderprob{\genDoc}{\pfDoc}$ to the \rendition
 probabilities of the  top $\numNormDocToDoc$ {\renderers}
\footnote{
In our experiments, we fixed $\numNormDocToDoc$ to a value guaranteed to be greater than
 $\spreadDocToDoc$ to handle cases where more than  $\spreadDocToDoc$
documents had high \rendition probabilities for $\query$.  However,
preliminary experiments indicated that choosing $\numNormDocToDoc =
\spreadDocToDoc$ did not substantially alter results. }
of $\pfDoc$.
Observe that in the first round, only the top \renderers of
the original query can receive non-zero scores.

Like all  pseudo-feedback techniques, both scoring methods
just presented can help ameliorate the crucial ``aspect recall''
problem.
We expect that the most improvement
relative to retrieval based directly on the original query  $\origQuery$
will occur when
the \pfDocsText contain effective
search terms not appearing in 
$\origQuery$,
and among the best \renderers of these \pfDocsText are relevant
documents containing the missing terms but not having high overlap
with $\origQuery$.  
Our third basic scoring method, \firstmention{\mcClusterToDoc}, makes
use of document clusters to take more explicit advantage of such
situations.  In particular, we choose \renderers of the
\pfDocsText from the set  $\clusters{\corpus}$ of document {\em clusters}
rather than from the set of documents.  Ideally, the best \renderers
would be clusters consisting only of relevant documents (thus increasing
recall), and some of these clusters  would represent
information whose importance is implied by
the query but not explicitly mentioned by it (thus increasing
``aspect recall'').  
In any event, integrating clusters into language-model-based retrieval
has recently been shown to yield substantial performance improvements
\cite{Kurland+Lee:04a,Liu+Croft:04a}.

While there are a huge number of clustering methods
to choose from, 
some quite sophisticated, we simply
create one cluster for every document by grouping together
that document's top $\sizeclust$
\renderers.
This method is convenient for us since we already need to compute
top \renderers; moreover,
it has proven effective in previous work, perhaps
because the highly overlapping clusters can
be seen as representing
different {\em facets} of the similarity structure of the corpus
\cite{Kurland+Lee:04a}.
Indeed, there is a history of successful applications of the general
nearest-neighbor approach
(e.g., \cite{Griffiths+Luckhurst+Willett:86a}).

Within each iteration,  \mcClusterToDoc scoring consists of two phases.  In the first, each cluster $\clust$ is credited
for every \pfDocText in its \repertoireText.
Note that in computing \repertoiresText, we chose to restrict the set of possible top
\renderers of 
a
 \pfDocText $\pfDoc$ to $\clusterswith{\pfDoc}$,
the set of clusters containing
$\pfDoc$: 
a cluster is only ``allowed'' to \render its constituent documents\footnote{In the
first iteration only, we treat the original query as a document
belonging to all clusters.}. (For the sake of readability, we suppress
this restriction in the \repertoireText notation below.) We thus have:
\begin{equation}
\label{eq:mcClusterToDoc}
\algScoreAtStepT{\abbrevMCclustToDoc}{\clust} \definedas 
\sum_{ \pfDoc \in \bestDocToClust}
\pfScore(\pfDoc)  \cdot 
\frac{\renderprob{\clust}{\pfDoc}}
{\normfactorDocToClust{\pfDoc}},
\end{equation}
where $\normfactorDocToClust{\pfDoc} = \sum_{\clust' \in
  \clusterswith{\pfDoc}} \renderprob{\clust'}{\pfDoc}$ re-scales
  $\clust$'s \rendition probability with respect to the set of
  clusters containing $\pfDoc$.

The purpose of the second phase of scoring is to convert the implicit
cluster ranking just computed into a document ranking, since the
output of each round should be a legal set of final retrieval results.
This is achieved by crediting each document for every cluster it is one
of the top $\spreadDocToClust$ \renderers of, with the restriction
(again suppressed in the \repertoireText notation below to enhance
readability) that a document may only \render a cluster that it
belongs to:
\begin{equation}
\label{eq:docToCluster}
\algScoreAtStepT{\abbrevMCclustToDoc}{\doc} \definedas
\sum_{\clust \in \bestClustToDoc}
\algScoreAtStepT{\abbrevMCclustToDoc}{\clust} \cdot 
\frac{\renderprob{\doc}{\clust}}
     {\normfactorClustToDoc{\clust}},
\end{equation}
where $\normfactorClustToDoc{\clust} = \sum_{\doc' \in \clust}
\renderprob{\doc'}{\clust}$ re-scales $\doc$'s \rendition probability
with respect to the set of documents within $\clust$.

\paragraph*{Remarks}
If the desired values of 
$\spreadDocToDoc$ and  $\spreadDocToClust$ (the parameters controlling
the 
sizes of the top-\renderer sets)
 and $\sizeclust$ (the parameter for
cluster size)  are known beforehand, then the clusters and
top-\renderer sets
can be computed off-line,
greatly reducing the amount of computation required at retrieval
time.  However, even if these parameters are 
not pre-specified,
 one can 
pre-compute
a ranking of all possible \renderers of each document;
this  still
results in significant computational savings at run-time.

We note that the iterative 
processes based upon the
latter two of the
basic scoring schemes we have just described 
can
be conceptualized
as a fixed-length random walk on a graph corresponding to a Markov chain whose structure
is determined by top-\renderer relationships.  In fact, the
\mcClusterToDoc scoring method is reminiscent of the term-to-document
Markov chain used by Lafferty and Zhai \cite{Lafferty+Zhai:01a} for
query expansion.  However, in our case it is not clear what insight is gained by
such a formulation; 
for instance,  how any stationary distribution should be interpreted
in the context of the retrieval task at hand is not obvious.

Finally, notice that
all three methods just outlined contain 
the standard
language-modeling approach \cite{Ponte+Croft:98a} as a 
degenerate
one-iteration  
(or half-iteration, for \mcClusterToDoc) 
case
where
$\spreadDocToDoc = |\corpus|$, and
for
\mcClusterToDoc, the cluster set
$\clusters{\corpus}$ corresponds to a
partition of $\corpus$ into single-element sets.

\newpage

\subsection{Coping with \overgeneralCAP}
\label{sec:iterate}

We have previously mentioned that one of our main interests is
increasing so-called ``aspect recall''.  Naturally, we want to
simultaneously retain high precision as well.  However, 
a potential drawback
of our iterative approach to \pfDocText processing is 
that engaging in multiple rounds threatens to exacerbate 
\overgeneralLow:
early contamination of the set of \pfDocsText 
with non-relevant documents
can seriously skew downstream \pfDocText sets away from the user's
true information needs.  Using clusters adds even more risk of
overgeneralization.  
We therefore propose a number of methods for addressing 
the \overgeneralLowH problem.
All
follow the same general strategy: ensure that information from the
original query 
$\origQuery$
plays a large role.

Two indirect techniques
all our algorithms employ
involve choosing appropriate values for
certain parameters.  
First, we limit
$\maxStepT$, the total number of rounds, to a relatively small number.
Second, since the initial iteration is the one that is
``closest'' to the original query, we give it privileged status by
considering the top $\pfNum$, rather than  $\spreadDocToDoc$,
\renderers of $\origQuery$.

We also consider a number of {\em re-scoring} techniques; these
directly use $\renderprob{\doc}{\origQuery}$ in some combination with
the output of one of the three basic scoring methods $\methodArb$
introduced above.  Recall that without re-scoring, the \pfDocText
\pfWeightsText $\pfScore(\doc)$ for round $\stepT+1$ would simply be the score
assigned by $\methodArb$ to document $\doc$ at the end of round
$\stepT$.

We borrow two
re-scoring techniques 
from research on cluster-based retrieval within
the language-modeling framework \cite{Kurland+Lee:04a,Liu+Croft:04a}.
Both affect only the output of the final round, since they were
introduced in the context of non-iterative methods.
The \firstmention{\mixtureAlg} 
technique
derives a new final
score
for each document $\doc$ by 
linear interpolation of $\scoreByMethodFinal{\doc}{\methodArb}$, 
$\doc$'s score in the final round  according to $\methodArb$,  
with 
the \rendition probability 
that $\doc$ assigns to 
the  original query
(after rescaling both quantities with 
respect to their maximum values
to 
ensure comparability):
\[
\mixtureParam \scoreByMethodFinal{\doc}{\methodArb} +
(1-\mixtureParam) \renderprob{\doc}{\origQuery}.
\]
It thus integrates
our iterated estimate of document relevance with surface
document-query similarity.

In contrast, the \firstmention{\rerank} method first discards all but
the top $\numretdocs$ documents according to
$\scoreByMethodFinal{\doc}{\methodArb}$; each remaining document
$\doc$ is then given the new score
$\renderprob{\doc}{\origQuery}$. 
Note that this method does not affect recall,
since
only the order of the retrieved results is changed.

Alternatively, we could alter the scores 
at the end of every round, rather than 
just  the final one, as an attempt to counteract \overgeneralLow
early in the process.  One idea, implemented in the
\firstmention{\boosting} technique, is to consider only the top
$\numretdocs$ documents in a given 
iteration or pass
to be likely to be informative \pfDocsText for the next round; the
scores of all the 
other documents are therefore zeroed.  The
\firstmention{\rerankBoosting} technique goes even further by
additionally
changing the scores of the
top $\numretdocs$ documents to $\renderprob{\doc}{\origQuery}$.  
That is, the \rerank technique is applied to each
round, rather than just to the final one.
Similarly, we can apply \mixtureAlg at each round, thus yielding the
\firstmention{\iterMixAlg}
technique.  
Note that this
method is more conservative than the original \mixtureAlg 
technique
because it
tends to prevent \pfDocsText with low surface similarity to the query
from being assigned high scores in early rounds.

\subsection{Estimating Rendition Probabilities}
\label{sec:lmspec}

\Rendition probabilities are the foundation upon which all our
algorithms are built.  To describe the method by which we estimate 
them, we first introduce some preliminary concepts.
Let $\altArbGroupMember$ be either a text string or a set of text
strings.  Denoting the number of times a term $\word$
occurs in 
$\altArbGroupMember$ by
$\freq{\word}{\altArbGroupMember}$,
for an $\wordseqlength$-term text sequence $\wordseq$ we define
$$\mlprob{\altArbGroupMember}{\wordseq} \definedas \prod_{\wordIndex=1}^{\wordseqlength}\frac{\freq{\word_\wordIndex}{\altArbGroupMember}}{\sum_{\word'}
\freq{\word'}{\altArbGroupMember}};$$
this is commonly known as $\altArbGroupMember$'s {\em maximum
likelihood estimate} (MLE) for the sequence.
The {\em Dirichlet-smoothed} 
version of the MLE is defined as
\begin{equation}
\dirichletLM{\altArbGroupMember}{\wordseq}{\dirichletParam} \definedas
\prod_{\wordIndex=1}^{\wordseqlength}\frac{\freq{\word_\wordIndex}{\altArbGroupMember} +
\dirichletParam\cdot\mlprob{\corpus}{\word_\wordIndex}}{\sum_{\word'}
\freq{\word'}{\altArbGroupMember} + \dirichletParam},
\label{eq:dirichlet}
\nonumber
\end{equation}
where the smoothing parameter $\dirichletParam$ controls the degree of
reliance on relative frequencies in the corpus rather than on the counts
in $\altArbGroupMember$.

While 
the Dirichlet-smoothed unigram language model just defined
has been used directly
\cite{Zhai+Lafferty:01a,Liu+Croft:04a},
we adopt the following variant:
for \renderer
$\rendererVar$ and text sequence $\wordseqvar$, we set
\begin{eqnarray*}
\renderprob{\rendererVar}{\wordseqvar} & \definedas &
\exp\left(-\kld{\bareMLProb{\wordseqvar}(\cdot)}{\bareDirProb{\rendererVar}(\cdot)}\right)
\\
& \propto & \sqrt[\vert \wordseqvar \vert]{\dirichletLM{\rendererVar}{\wordseqvar}{\dirichletParam}},
\end{eqnarray*}
where $D$ is the 
KL divergence, 
which has formed the basis for other
ranking principles as well
\cite{Xu+Croft:99a,Lafferty+Zhai:01a,Ng:00a,Kurland+Lee:04a};
the two arguments to $D$ are treated as
distributions over terms rather than term sequences;
and 
the omitted
factor 
(which is of independent interest in other contexts \cite{Kurland+Lee:05a})
drops out in the re-scaling performed by the \mcDocToDoc and
\mcClusterToDoc scoring methods.
Our formulation provides some mathematical justification for Lavrenko
et al.'s ``heuristic adjustment'' \cite{Lavrenko+al:02a}, proposed 
to handle underflow problems in processing long documents, to take 
the {\em
geometric mean}
of 
$\dirichletLM{\rendererVar}{\wordseqvar}{\dirichletParam}$
rather than
$\dirichletLM{\rendererVar}{\wordseqvar}{\dirichletParam}$ itself.

\section{Related Work} 
\label{sec:relwork} 

The fundamental principle underlying pseudo-feedback-based 
methods is that the top-ranked documents retrieved in response to a
query may contain additional information regarding the user's
information need.  
The canonical approach is to treat the pseudo-feedback documents as if
they had actually been deemed relevant by the user, and then apply
relevance-feedback techniques \cite{Ruthven+Lalmas:03a} to them.
One such line of
work is to use the feedback documents to re-weight query terms and/or
to identify additional terms with which to augment the query; since
our wholesale replacement of the query with pseudo-queries can be
considered an extension of this idea, in Section \ref{sec:exp} we
compare against one well-known instantiation, namely, Rocchio
\cite{Rocchio:71a}.  
Within the language-modeling retrieval framework,
treating the feedback documents as relevant often means estimating
\rendition probabilities using the feedback pool (where the members
may be differentially weighted) as data 
\cite{Lavrenko+Croft:01a,Lavrenko:02a,Lafferty+Zhai:01a,Zhai+Lafferty:01b,Tao+Zhai:04a}.

Lafferty and Zhai \cite{Lafferty+Zhai:01a} proposed an
iterative probability-estimation sub-routine that alternates between
terms and documents, which is remiscent of the shifting between
clusters and documents that our \mcClusterToDoc algorithm represents;
but their intended application is not a direct scoring of potential
retrieval candidates.  Lavrenko and Croft's 
{\em \relModelText}
algorithm \cite{Lavrenko+Croft:01a} 
has the same goal as ours, is also based on language models,  and
posts state-of-the-art performance; Section \ref{sec:exp} describes it
in more detail and reports the results of our experimental comparisons
against it.

\overgeneralCap has long been recognized as a 
key concern for 
pseudo-feedback approaches
\cite{Mitra+Singhal+Buckley:98a,Croft+Harper:79a}, and 
hence a number
of coping techniques have been previously introduced.
One example is the application of boolean filters and 
term co-occurrence analysis \cite{Mitra+Singhal+Buckley:98a}.
The techniques we adopted focus instead on incorporating \rendition
probabilities for the original query,
borrowing from previous work \cite{Zhai+Lafferty:01b,Liu+Croft:04a,Kurland+Lee:04a}.

\section{Experiments}
\label{sec:exp}

\newcommand{\basicTableSize}{}
\newcommand{\basicTableCaption}{Comparison against the baseline.
Statistically significant differences with the baseline are 
marked with a star (*).
Bold: best performance for each setting (column). 
Italics: results superior to the baseline.  
}

\newcommand{\better}[1]{\mathit{#1}}

\begin{table*}
\begin{flushleft}
\centering
\begin{tabular}{|l|c|c|c|c|c|c|}
\hline
& \multicolumn{2}{|c|}{AP89}& \multicolumn{2}{|c|}{AP88+89}& \multicolumn{2}{|c|}{LA+FR}\\ \cline{2-7}
& {prec} & {recall} & {prec} & {recall} & {prec} & {recall} \\ \hline
Baseline& $20.74\%^{~}$& $48.67\%^{~}$& $24.26\%^{~}$& $66.62\%^{~}$& $21.72\%^{~}$& $48.81\%^{~}$\\ \hline
\hline
\abbrevItFifo& $\better{23.12\%}^{*}$& $\better{55.20\%}^{~}$& $\better{28.28\%}^{*}$& $63.68\%^{~}$& $\better{22.07\%}^{~}$& $47.81\%^{~}$\\ \hline
\abbrevMCdocToDoc& $\better{22.81\%}^{*}$& $\better{57.34\%}^{~}$& $\better{28.27\%}^{*}$& $\better{71.13\%}^{~}$& $\better{22.48\%}^{*}$& \mbox{\boldmath$59.67\%^{~}$}\\ \hline
\abbrevMixMCdocToDoc& $\better{24.48\%}^{*}$& $\better{59.77\%}^{*}$& $\better{30.28\%}^{*}$& $\better{75.57\%}^{*}$& $\better{22.98\%}^{*}$& $\better{56.00\%}^{*}$\\ \hline
\abbrevMCclustToDoc& $\better{23.43\%}^{*}$& \mbox{\boldmath$63.57\%^{*}$}& $\better{28.76\%}^{*}$& \mbox{\boldmath$80.15\%^{*}$}& $\better{23.24\%}^{*}$& $\better{56.79\%}^{~}$\\ \hline
\abbrevMixMCclustToDoc& \mbox{\boldmath$24.56\%^{*}$}& $\better{62.77\%}^{*}$& \mbox{\boldmath$31.09\%^{*}$}& $\better{76.15\%}^{*}$& \mbox{\boldmath$23.40\%^{~}$}& $\better{54.57\%}^{*}$\\ \hline
\end{tabular}
\caption{\label{tab:BasicTable} \basicTableCaption} 

\end{flushleft}
\end{table*}

\newcommand{\relModelTableSize}{\small}
\newcommand{\relModelTableCaption}{Comparison 
of the best \pfDocText algorithms
against Rocchio, 
the
\relModelText, and the \clipped \relModelText. Statistically significant
differences with these algorithms are marked with 
$\rocDiff$, $\relDiff$, and $\clippedDiff$, 
 respectively. Bold: best performance for each setting (column). }

\begin{table*}
\relModelTableSize
\hspace*{.77in}
\begin{tabular}{|l|c|c|c|c|c|c|}
\hline
& \multicolumn{2}{|c|}{AP89}& \multicolumn{2}{|c|}{AP88+89}& \multicolumn{2}{|c|}{LA+FR}\\ \cline{2-7}
& {prec} & {recall} & {prec} & {recall} & {prec} & {recall} \\ \hline
\abbrevMCclustToDoc& $ 23.43 \%^{~~} $& \mbox{\boldmath$ 63.57 \% $}$^{~~}$& $ 28.76 \%^{\relDiff\clippedDiff} $& $ 80.15 \%^{\relDiff\clippedDiff} $& $ 23.24 \%^{\rocDiff~} $& \mbox{\boldmath$ 56.79 \% $}$^{\relDiff~}$\\ \hline
\abbrevMixMCclustToDoc& $ 24.56 \%^{\rocDiff~} $& $ 62.77 \%^{~~} $& $ 31.09 \%^{~~} $& $ 76.15 \%^{\relDiff\clippedDiff} $& \mbox{\boldmath$ 23.40 \%$}$^{\rocDiff~} $& $ 54.57 \%^{\rocDiff\relDiff} $\\ \hline
\hline
Rocchio& $ 22.85 \%^{~~} $& $ 58.23 \%^{~~} $& $ 30.69 \%^{~~} $& $ 76.02 \%^{~~} $& $ 18.21 \%^{~~} $& $ 52.26 \%^{~~} $\\ \hline
\abbrevRelModel& $ 24.72 \%^{~~} $& $ 58.08 \%^{~~} $& \mbox{\boldmath$ 32.72 \%$}$^{~~} $& $ 81.60 \%^{~~} $& $ 22.03 \%^{~~} $& $ 46.59 \%^{~~} $\\ \hline
\abbrevRelModelClipped& \mbox{\boldmath$ 26.17 \%$}$^{~~} $& $ 63.48 \%^{~~} $& $ 32.51 \%^{~~} $& \mbox{\boldmath$ 82.10 \% $}$^{~~}$& $ 22.34 \%^{~~} $& $ 56.65 \%^{~~} $\\ \hline
\end{tabular}
\caption{\label{tab:RelModelComp} \relModelTableCaption} 

\end{table*}

To examine
the effectiveness of our 
algorithms
 and to determine how much
various aspects of our proposed retrieval framework contribute, we designed a
number of evaluation experiments.  

First,  we compare
the performance of our algorithms to 
that of 
a 
language-model-based approach (henceforth {\em baseline})
in which documents are ranked
according to $\renderprob{\doc}{\origQuery}$.  
This comparison serves not only to see whether our methods can outperform
an effective retrieval system, but also to
highlight the merits (or lack thereof) of engaging in multiple
iterations of \pfDocText processing, since, as noted above, 
conceptually
the
basic language-modeling approach corresponds to a single round or pass of our algorithms.

We also test
our techniques for \pfDocText-processing  against 
a well known and highly effective pseudo-feedback method, 
the Rocchio algorithm \cite{Rocchio:71a} as applied to top-retrieved
documents.

Finally, we study whether our particular ways of utilizing language
models
are beneficial by testing how well they perform against the {\em
\relModelText} \cite{Lavrenko+Croft:01a,Lavrenko+Croft:03a}.
The latter approach takes a generative perspective: assuming that
there is a single {\em relevance} language model $\relModel$
underlying the creation of both $\origQuery$ and the documents
relevant to $\origQuery$, documents are ranked by their degree of
``match'' with $\relModel$, rather than
by
 how well they directly match the query
or set of \pfDocsText.  In implementation, Lavrenko and Croft
estimate $\relModel$ by combining the language models of those
documents assigning the highest \rendition probabilities to
$\origQuery$.  
Thus, the relevance-model, similarly to our algorithms, 
is a pseudo-feedback-based language-modeling approach,
but clearly the specific way in which
document-based language models are used is quite different from the
ways our algorithms employ them, and Lavrenko and Croft made no
explicit mention of clusters.

Although our reference comparison models operate in different spaces
(vector space vs. 
the probability simplex), in \cite{Lavrenko+Croft:03a}
it is observed that  if {\em i.i.d sampling}
sampling is used for \relModelText estimation, then both the
pseudo-feedback version of Rocchio and the \relModelText 
 utilize a 
linear combination of 
the
top-retrieved documents' models 
to construct an {\it expanded query model}.

We 
conducted our experiments on the following three corpora, drawn from TREC data:
\begin{center}
\begin{tabular}{|lrcc|}\hline
corpus & \# of docs & queries &\# of relevant docs \\ \hline
AP89 & 84,678 & 1-46,48-50 & 3261 \\
AP88+89 & 164,597 & 101-150 & 4805\\
LA+FR  & 187,526 & 401-450 & 1391 \\  \hline
\end{tabular}
\end{center}
The AP89 corpus was 
pre-processed with
the Porter stemmer. For AP88+89 the Krovetz
stemmer was used, and 
both INQUERY stopwords
\cite{Allan+al:00a} and length-one tokens were removed 
to comply with the processing policy in \cite{Lavrenko+Croft:01a}.
For LA+FR, 
which is 
part of the TREC-8 corpus, neither 
stemming nor stopword removal was applied. 
It is relatively heterogeneous, and the
LA dataset with TREC8 queries is considered to be 
difficult 
\cite{Hu+Bandhakavi+Zhai:03a}.

For queries, we used 
the titles of TREC topics 
rather than the full
descriptions, resulting in short queries containing 2-5 terms on
average.

We use both average non-interpolated precision and recall at
$\numretdocs=1000$ as our evaluation measures. Statistically
significant differences in performance are determined using the
two-sided Wilcoxon test at the $95\%$ confidence level.

\subsection{Implementation}
\label{sec:expSetup}

We employed the Lemur toolkit \cite{Ogilvie+Callan:2001} for a number
of our experiments.  To collect pseudo-feedback, we 
used $\renderprob{\doc}{\query}$ to create an initial ranking
All parameters were set to values optimizing average non-interpolated precision.

Our implementation of Rocchio used the vector-space model with log
tf.idf term weighting to represent queries and documents.  Similarity
was measured via the inner product.  The free parameters were:
(i) $\pfNum$ - the number of top-retrieved documents used for feedback, (ii) the number of terms to augment
the original query with, and (iii) the weighting coefficient for the augmenting terms
(we only used positive feedback).
Note that while the number of (augmenting) terms is not modeled in
Rocchio's original method, we varied it to obtain better performance and comply
with the optimization steps we implemented for the \relModelText
(details further below).
Our implementation  yielded accuracies consistent with 
previously reported (optimized) results.

Our Lemur-based implementation of the \relModelText
utilized {i.i.d sampling} (following \cite{Lavrenko+Croft:03a}) to
construct $\relModel$; 
the divergence $\kld{\relModel}{\renderprob{\doc}{\cdot}}$ served
as ranking criterion.
The free parameters were
$\pfNum$ (the number of top-retrieved documents)  and an interpolation parameter controlling
the evaluation of 
the
top retrieved documents' language models.

We also experimented with 
{\em \clipping} 
the \relModelText \cite{Connell+al:04a,Cronen-Townsend+Zhou+Croft:04a}
to assign non-zero probability to only a restricted number of terms
(up to a maximum of several hundred).
This 
modification can be viewed as regulating the degree of
query expansion, or as an efficiency-improving heuristic.

Some of our algorithms have quite a few free parameters.  To 
help prevent  our algorithms 
from enjoying an 
unfair advantage 
due to this
fact alone, we implemented the
following policies.
The language models forming the basis of 
both
our methods and the baseline had the Dirichlet
smoothing parameter value fixed at
$\dirichletParam$ $=2000$, following \cite{Zhai+Lafferty:01a}.
All
parameters 
shared by our methods
and the \relModelText{s} (e.g., $\pfNum$)
were set identically for all the algorithms,
with the following  search ranges: \\
\parmark $\spreadDocToDoc$, the number of top \renderers  considered: $\{5,10,20,\ldots,100\}$ for
\fifoAlg; $\{5,10,20,30,40\}$ for \mcDocToDoc; from 
$\{1,2,3,4\}$ for
\mcClusterToDoc. \\
\parmark $\pfNum$, the  number of best \renderers retrieved at the
first iteration: $\set{5} \cup \set{10, 20, ..., 100} \cup \set{200,
  300, 400, 500}$. \\
\parmark $\spreadDocToClust$, the number of documents to which a
cluster's score is distributed (Equation~\ref{eq:docToCluster}):
\set{5,10,20,30,40} for AP89 and AP88+89; \set{5,10} for LA+FR.\\
\parmark $\sizeclust$, cluster size: 40 for AP89 and AP88+89; 10 for
LA+FR. \\
\parmark $\maxStepT$, the number of rounds: 1--2, \mcClusterToDoc;
1--5, \fifoAlg and
\mcDocToDoc.

\subsection{Main results}
\label{sec:expResults}

\newcommand{\rerankTableSize}{}
\newcommand{\rerankTableCaption}{Comparison of techniques for
\overgeneralLow prevention, the \mcDocToDoc scoring method.  Bold:
best performance for a given evaluation setting (column). Note that
\abbrevReRank ~improves recall over the basic algorithm
(\abbrevNoPrev) as it achieves optimal precision with a different
parameter setting.}
\begin{table*}
\begin{flushleft}\rerankTableSize
\centering
\begin{tabular}{|l|c|c|c|c|c|c|}
\hline
& \multicolumn{2}{|c|}{AP89}& \multicolumn{2}{|c|}{AP88+89}& \multicolumn{2}{|c|}{LA+FR}\\ \cline{2-7}
& {prec} & {recall} & {prec} & {recall} & {prec} & {recall} \\ \hline
\abbrevNoPrev& $ 22.83 \% $& $ 51.43 \% $& $ 28.49 \% $& $ 69.99 \% $& $ 21.55 \% $& $ 53.20 \% $\\ \hline
\abbrevReRank& $ 22.81 \% $& $ 57.34 \% $& $ 28.27 \% $& $ 71.13 \% $& $ 22.48 \% $& \mbox{\boldmath$ 59.67 \% $}\\ \hline
\abbrevIntRer& \mbox{\boldmath$ 24.48 \% $}& \mbox{\boldmath$ 59.77 \% $}& \mbox{\boldmath$ 30.28 \% $}& \mbox{\boldmath$ 75.57 \% $}& \mbox{\boldmath$ 22.98 \% $}& $ 56.00 \% $\\ \hline
\abbrevBoosting& $ 19.71 \% $& $ 51.09 \% $& $ 26.86 \% $& $ 63.37 \% $& $ 16.84 \% $& $ 31.63 \% $\\ \hline
\abbrevRerankBoosting& $ 22.76 \% $& $ 58.02 \% $& $ 27.96 \% $& $ 68.05 \% $& $ 22.49 \% $& $ 59.45 \% $\\ \hline
\abbrevIterIntRer& $ 24.27 \% $& $ 59.09 \% $& \mbox{\boldmath$ 30.28 \% $}& \mbox{\boldmath$ 75.57 \% $}& \mbox{\boldmath$ 22.98 \% $}& $ 56.00 \% $\\ \hline
\end{tabular}
\caption{\label{tab:RerankBoost} \rerankTableCaption} 

\end{flushleft}
\end{table*}

We report results using the following abbreviations.
\begin{center}
\begin{tabular}{ll}
\abbrevItFifo &  \fifoAlg \\
\abbrevMCdocToDoc &\mcDocToDoc \\
\abbrevMCclustToDoc & \mcClusterToDoc \\
\end{tabular}
\end{center}
The prefix ``\abbrevMixPrefix'' indicates that the (non-iterated)
\mixtureAlg technique was employed for coping with \overgeneralLow; in
all other cases, \rerank was applied. 
Results for other query-drift amelioration mechanisms are reported
later in this section.

Table \ref{tab:BasicTable} 
compares our
algorithms' performance to that of the language-model
baseline.
We see
that 
almost all
our 
 methods outperform
the language model
 in both 
 average precision and
recall, often to a statistically significant degree.
Moreover, it is clear that our cluster-based algorithm \mcClusterToDoc,
in either its \rerank or \mixtureAlg version, yields the best results,
outperforming not only the
language-modeling approach but all the non-cluster-based
algorithms we have proposed. This finding further reinforces
conclusions previously drawn in the literature regarding the
advantages of using clusters to represent cross-document contextual
information \cite{Kurland+Lee:04a,Liu+Croft:04a}; and the fact that we
see especially large improvements in recall provides partial support
towards our hypothesis that clusters can potentially alleviate the
problem of aspect recall.

Moving on to our second main comparison, Table~\ref{tab:RelModelComp}
shows that our cluster-based methods 
usually 
yield 
results
that are better (sometimes to a significant degree) than those of
Rocchio. Comparing our algorithms' performance to that of the
two versions of the \relModelText, we observe the following: on AP89,
our cluster-based methods yield results that are,
in a statistical sense,
indistinguishable from those of the (\clipped) \relModelText; on LA+FR
our methods tend to be superior to the (\clipped) \relModelText (sometimes
significantly
so); but on AP88+89, the
(\clipped) \relModelText generally 
performs 
significantly better than
our cluster-based methods.
In interpreting these results, though, it is crucial to note that
(1) our implementation of the
(\clipped) \relModelText involved an extremely wide-ranging search over the
parameter space, whereas
as discussed in Section~\ref{sec:expSetup},
our 
methods 
only explored
moderate parameter-setting ranges;
and (2) 
clipping is a heuristic that could
potentially be adapted for use by our document- or cluster-based
language models.

While
some preliminary results 
indicate that the performance of our methods can be further improved 
by 
more exhaustive parameter tuning, we believe that the main message of
Table~\ref{tab:RelModelComp} is that we can achieve performance
competitive with optimized state-of-the-art pseudo-feedback methods
with relatively little optimization effort.

\medskip \subsection{Further analysis}
Examination of the average-precision results in Table
\ref{tab:BasicTable} reveals that the \mixtureAlg technique is usually
more effective at coping with \overgeneralLow than \rerank.
Table~\ref{tab:RerankBoost} provides a more extensive comparison of
the full set of \overgeneralLowH-prevention techniques we have
proposed.  For simplicity, we report only the results of applying
these methods in conjunction with the \mcDocToDoc algorithm.  It is
apparent that most of our techniques achieve comparable 
or better 
precision than is obtained by the original method by itself,
and that \mixtureAlg is
the ``winner'', 
although the more conservative iterated version (\iterMixAlg) ties it
on two corpora.  Investigation into the
optimal parameter settings revealed that, while performance for
\mcDocToDoc itself was optimized at a low number of iterations (which
would have the effect of keeping precision from degrading),
performance when prevention techniques were applied was best for a
larger number of iterations, enabling increase in recall along with
preservation of high precision rates.

\newcommand{\vtrwidth}{2.4in}
\begin{figure*}[t]
\hspace*{-.35in}
\begin{tabular}{ccc}
{\bf AP89} & {\bf AP88+89} & {\bf LA+FR} \\
\epsfig{file= 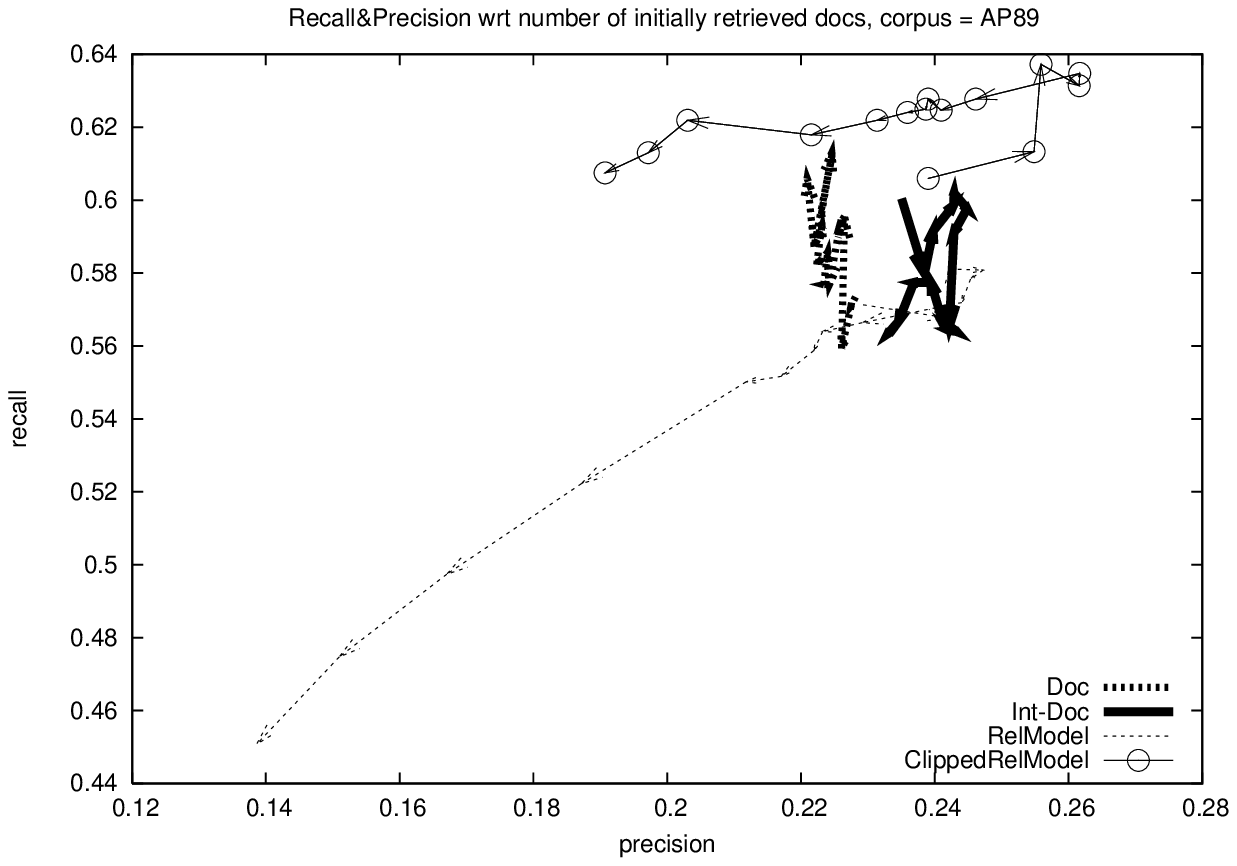,width=\vtrwidth}
&
\epsfig{file= 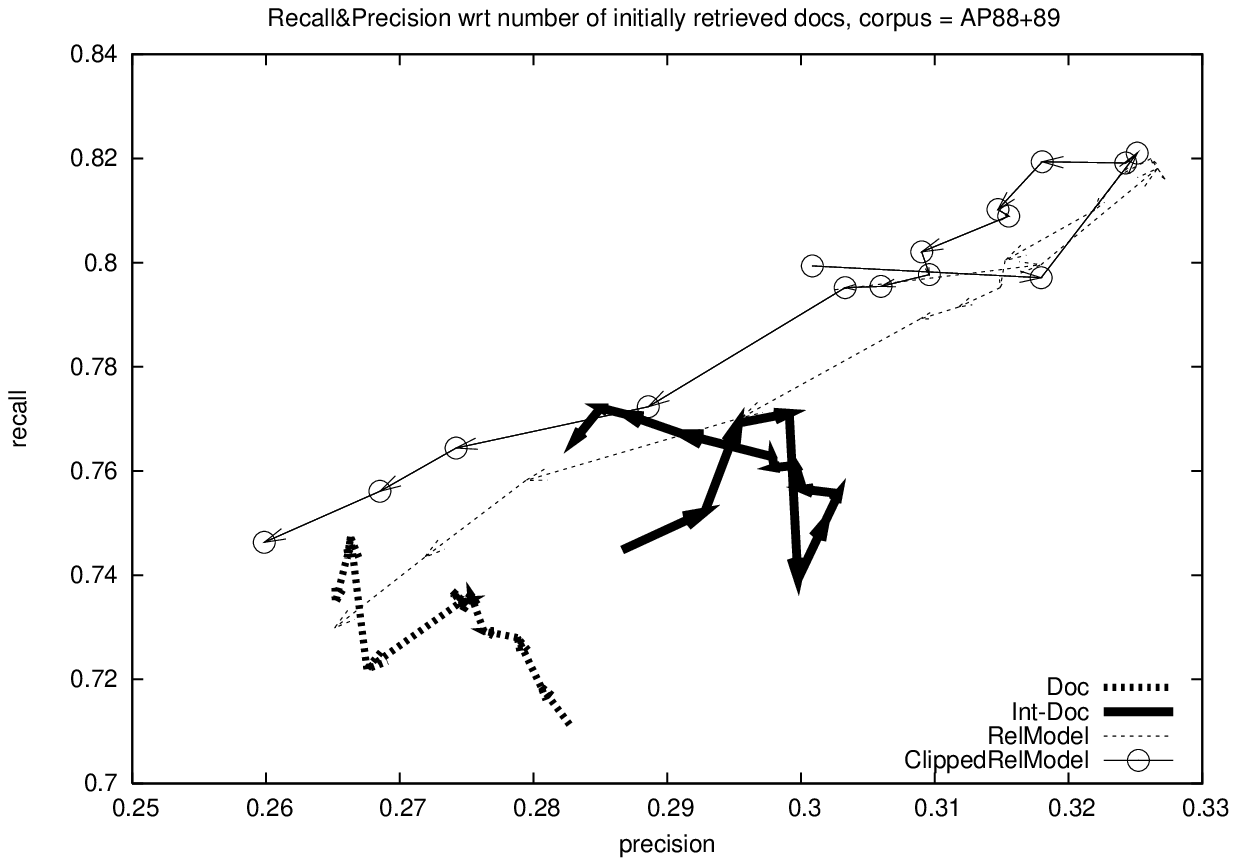,width=\vtrwidth}
&
\epsfig{file= 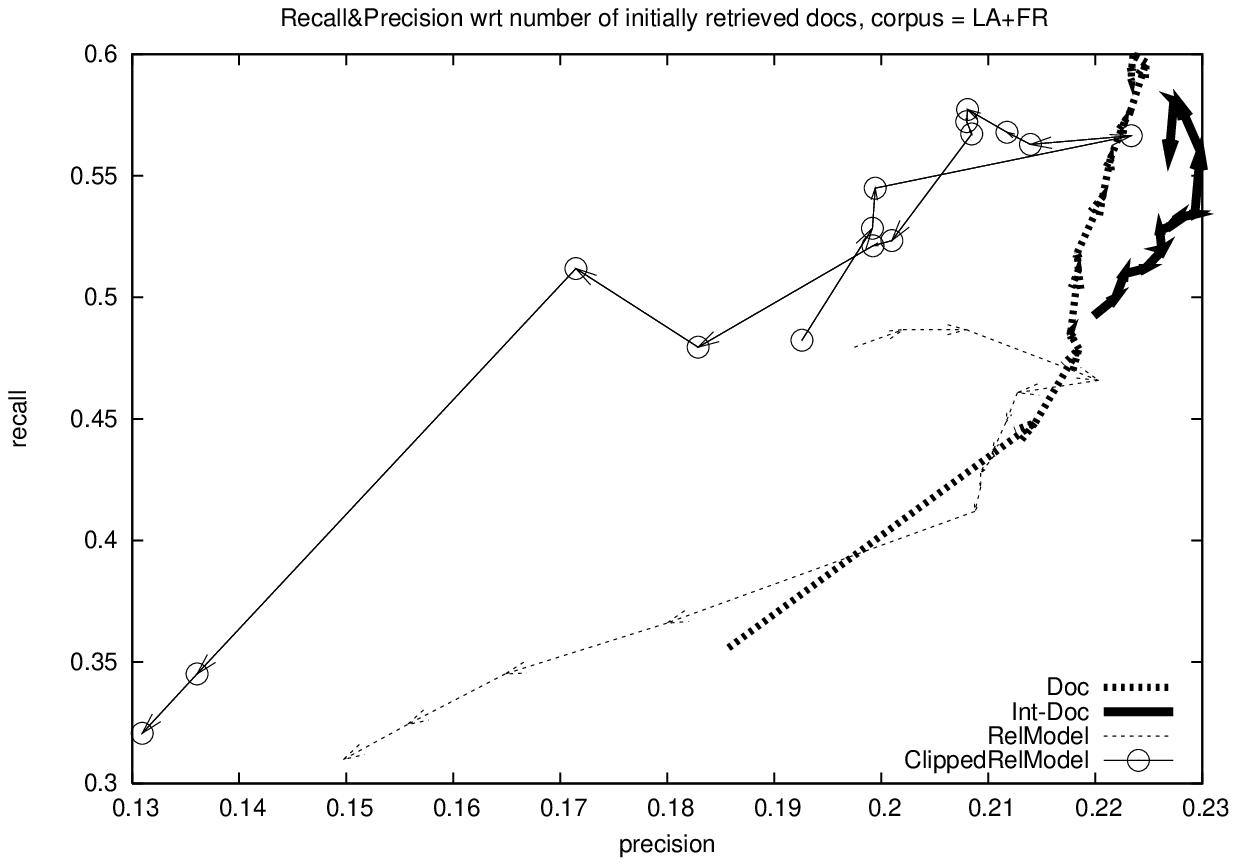,width=\vtrwidth} \\
\epsfig{file= 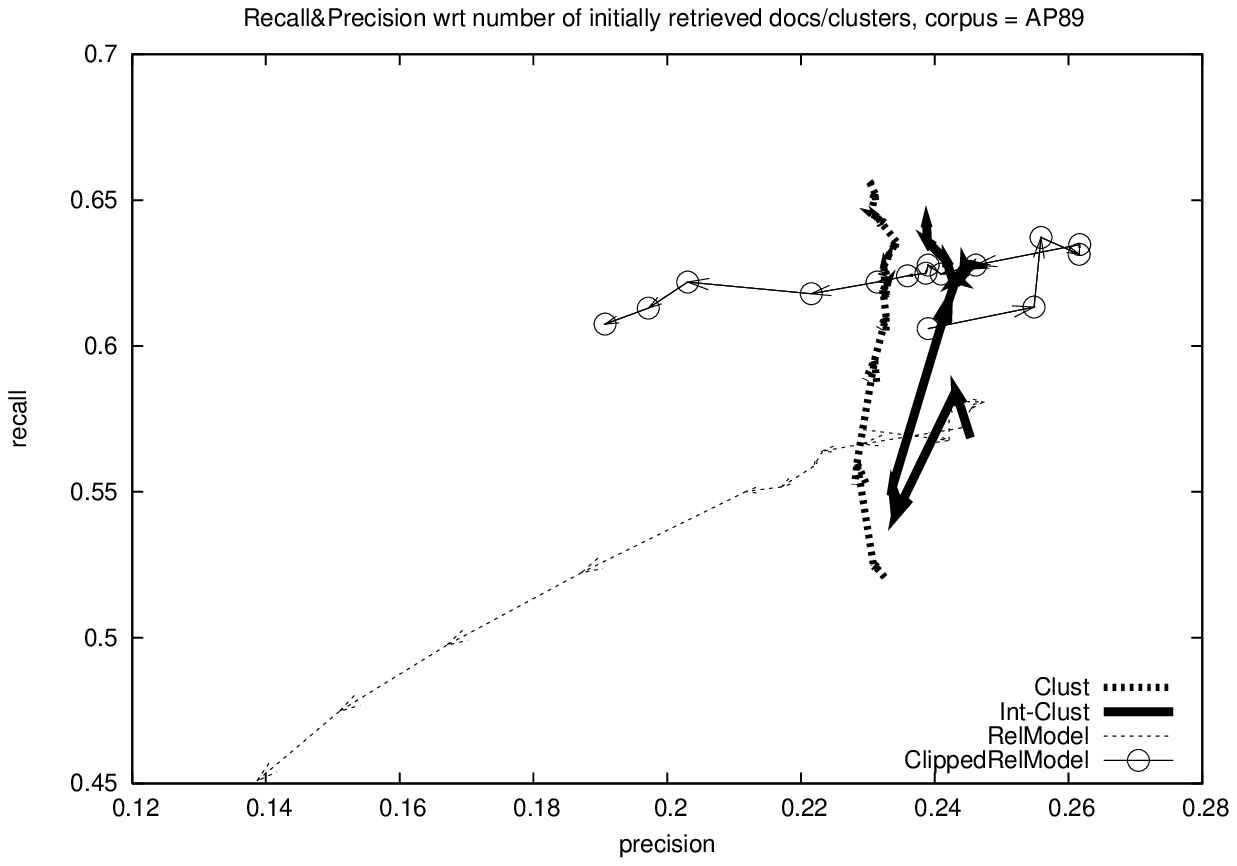,width=\vtrwidth}
&
\epsfig{file= 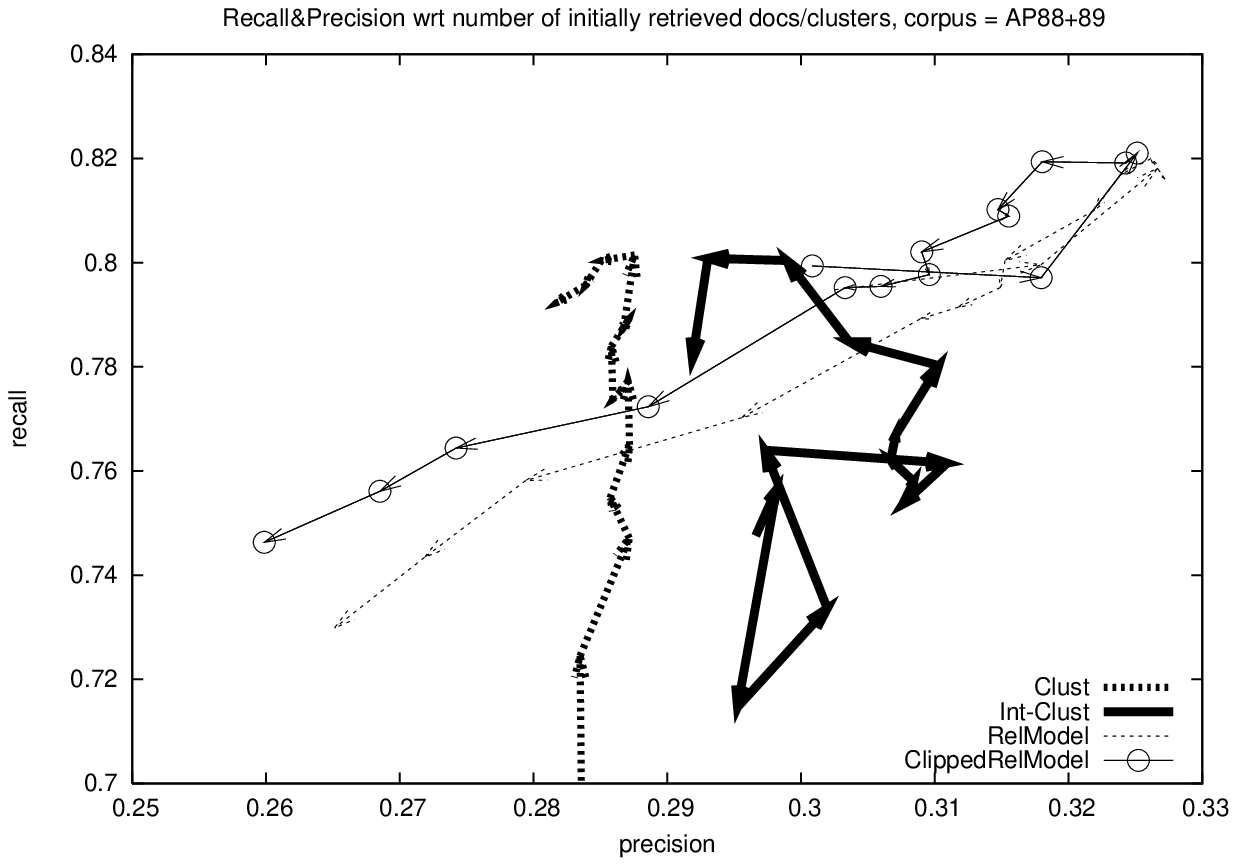,width=\vtrwidth}
&
\epsfig{file= 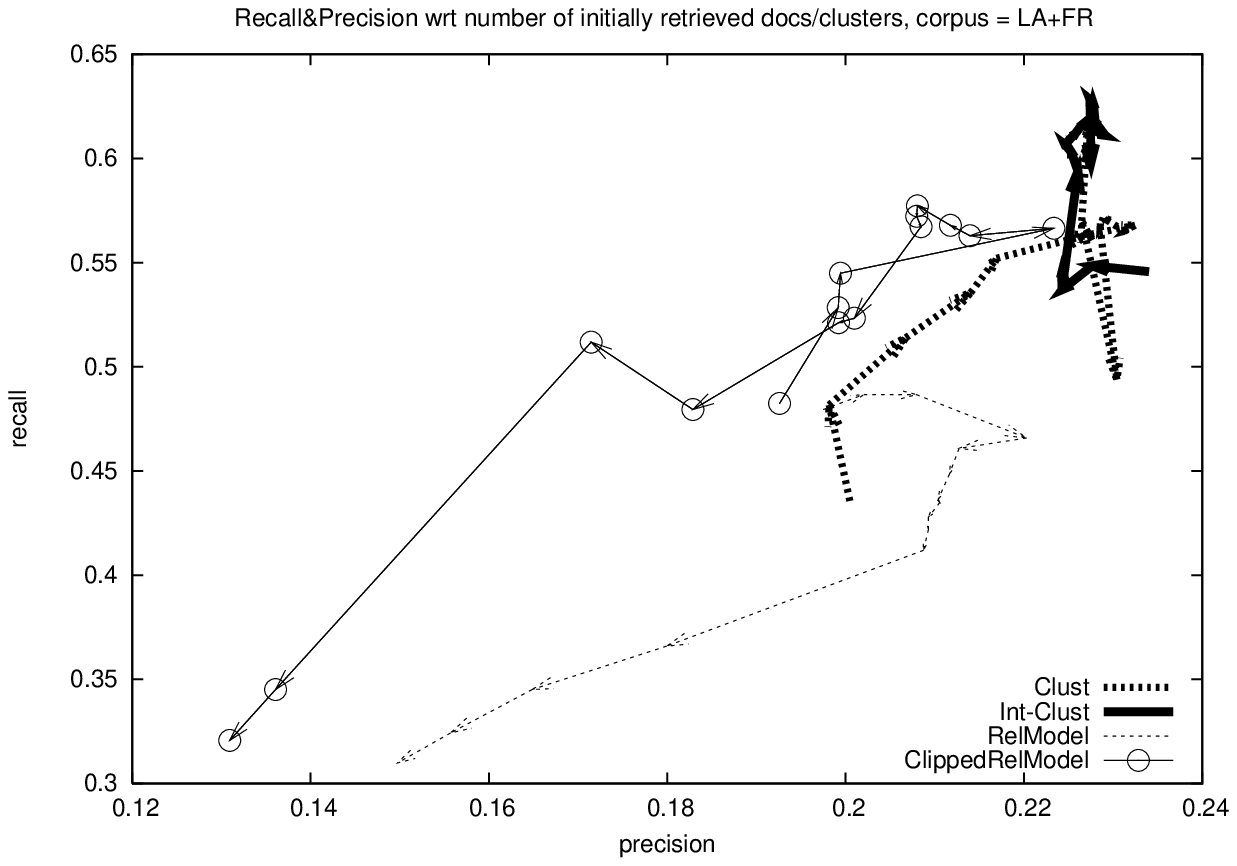,width=\vtrwidth}\end{tabular}
\caption{\label{fig:topRetVary} Average precision vs. recall as
  $\pfNum$, the number of
initially retrieved best \renderers for the original query, takes the values
5,10,20,...100,200,...,500.  Top and bottom rows show the best non-cluster-
and cluster-based algorithms, respectively, against the 
(\clipped)
\relModelText (note the relatively severe degradation in precision
--- ``leftwards motion'' ---
of the latter).
To show detail, the 
plots are not to the same scale.}
\end{figure*} 

Finally, we 
explored
a well-known weakness
 of 
language-model-based
approaches using pseudo-feedback:
sensitivity to the number $\pfNum$ of 
documents 
initially
retrieved \cite{Tao+Zhai:04b}. 
First, we see from Figure \ref{fig:topRetVary} that the precision of
our novel algorithms is much less affected by increases in
$\pfNum$ than the precision of the
(\clipped) \relModelText, which indicates the merits of
both of our \overgeneralLowH prevention techniques (although clearly
\mixtureAlg almost always outperforms \rerank at all values of  $\pfNum$).
It is also interesting to observe that increasing  $\pfNum$ tends to
have a positive influence on the recall of our cluster-based methods
(which, 
after all, were posited to improve aspect recall), whereas eventually it has a
negligible or negative influence on the 
(\clipped) \relModelText's
recall.
In short, the performance curves for our algorithms tend to move
vertically, whereas the 
(\clipped) \relModelText's curves seem to exhibit
more horizontal movement.
These trends 
suggest that our methods and the \relModelText have
complementary strengths.

\section{Conclusions}

We presented a novel iterative pseudo-feedback
approach to ad hoc information retrieval using 
cluster-based 
language
models. Starting from the 
original query, 
our methods 
repeatedly
seek 
potentially good \renderers of 
a current set of \pfDocsText,
guided by
the hypothesis that documents that are the best \renderers of a \pfDocText
may be good alternate \renditions of it.

One of the major challenges facing today's retrieval engines is
the problem of ``aspect recall''. To alleviate this problem,  we
proposed to 
take advantage of
corpus structure via
the consideration of
cluster-based language models as potential \renderers; the key idea is
that clusters can serve as a rich source of information regarding
corpus aspects.
Likewise,
we examined 
several techniques for 
reducing {\em \overgeneralLow}, which is yet 
another obstacle that both traditional and language-modeling-based
pseudo-feedback approaches need to overcome. 
As evidence that our techniques are effective, we saw that
our algorithms showed significant improvements in performance with
respect to a 
standard
language-modeling approach, and produced
results rivaling 
those of
other
state-of-the-art pseudo-feedback methods. 

For 
future work, we 
plan to
look into analyzing our methods in  real-feedback
settings,
e.g., \cite{Soboroff+Robertson:03a,
Aalbersberg:92a, Allan:03a}.
Furthermore, we would like to incorporate and examine
additional clustering approaches for modeling corpus structure.

\paragraph*{Acknowledgments}
We thank David Fisher for 
technical assistance with Lemur, and Jon
Kleinberg and the anonymous reviewers for valuable discussions and comments.
We also thank CMU for its hospitality during the year.
This paper is based upon work supported in part by the National
Science Foundation (NSF) under grant no.  IIS-0329064 and CCR-0122581;
SRI International under subcontract no. 03-000211 on their project
funded by the Department of the Interior's National Business Center;
and by an Alfred P. Sloan Research Fellowship. Any opinions, findings,
and conclusions or recommendations expressed are those of the authors
and do not necessarily reflect the views or official policies, either
expressed or implied, of any sponsoring institutions, the
U.S. government, or any other entity.

\end{document}